\documentclass[english,aps,showpacs, showkeys, notitlepage]{revtex4-2}
\usepackage[T1]{fontenc}
\usepackage[latin9]{inputenc}
\usepackage{color}
\usepackage{babel}
\usepackage{amsmath}
\usepackage{graphicx}
\usepackage[unicode=true,pdfusetitle,
 bookmarks=true,bookmarksnumbered=false,bookmarksopen=false,
 breaklinks=false,pdfborder={0 0 0},pdfborderstyle={},backref=false,colorlinks=true]
 {hyperref}
\begin{document}
\title{Cosmology of unimodular Born-Infeld-$f\left(R\right)$ gravity}
\author{Salih Kibaro\u{g}lu$^{1}$}
\email{salihkibaroglu@maltepe.edu.tr}

\author{Sergei D. Odintsov$^{2,3,5}$}
\email{odintsov@ice.csic.es}

\author{Tanmoy Paul$^{4,5}$}
\email{tanmoy.paul@visva-bharati.ac.in}

\date{\today}
\begin{abstract}
We propose a novel modified gravity: unimodular generalization of
the Born-Infeld-$f(R)$ gravity within the framework of cosmology.
After formulating the action corresponding to the generalized Born-Infeld-$f(R)$
gravity, we present a reconstruction scheme of this unimodular extension
to achieve various cosmological eras of the universe. Interestingly,
the unimodular generalization of Born-Infeld-$f(R)$ gravity proves
to be suitable for inflation (de-Sitter and quasi de-Sitter) along
with power law cosmology which actually includes from reheating to
the Standard Big-Bang cosmology. Apart from the inflation, such generalized
Born-Infeld-$f(R)$ theory is also capable to trigger non-singular
bouncing cosmology for suitable forms of the unimodular Lagrange multiplier,
which are exotic for the standard Einstein-Hilbert gravity and cannot
be realized in that case. The possible consequences are discussed. 
\end{abstract}
\affiliation{$^{1)}$Department of Basic Sciences, Faculty of Engineering and Natural
Sciences, Maltepe University, 34857, Istanbul, Turkey}
\affiliation{$^{2)}$Institute of Space Sciences (ICE, CSIC) C. Can Magrans s/n,
08193 Barcelona, Spain}
\affiliation{$^{3)}$ICREA, Passeig Luis Companys, 23, 08010 Barcelona, Spain}
\affiliation{$^{4)}$Department of Physics, Visva-Bharati University, Santiniketan
731235}
\affiliation{$^{5)}$Labaratory for Theoretical Cosmology, International Centre
of Gravity and Cosmos, Tomsk State University of Control Systems and
Radioelectronics (TUSUR), 634050 Tomsk, Russia}
\maketitle

\section{Introduction}

Currently, we are going through a cosmological era where, on one hand,
we have an ample amount of cosmological data like the spectral tilt
and the amplitude of primordial curvature perturbation, the tensor-to-scalar
ratio that describes the early phase of the universe; and the dark
energy EoS parameter, the Om(z) parameter which are related to the
late time dynamics of the universe. However, on the other hand, the
theoretical cosmology is still riddled with some serious issues like
--- why does the universe pass through acceleration at high and low
curvature regimes? Did the universe begin through a curvature singularity
or through a non-singular bounce? etc. Regarding the early universe
scenario, inflation earned the most attention as it successfully solves
the horizon and flatness problems, and also, the primordial curvature
perturbation generated from inflation turns out to be nearly scale-invariant
that is indeed consistent with the Planck observations \citep{Guth:1980zm,Linde:1990flp,Langlois:2004de,Riotto:2002yw,Baumann:2009ds}.
However, extrapolating backwards in time, inflation is plagued with
a curvature singularity (known as Big Bang singularity) due to the
geodesic incompleteness. Maybe, the Big Bang singularity is just a
limitation of the classical theory of gravity which fails to describe
the universe at such small scales. A yet-to-be-built quantum theory
of gravity may rescue such curvature singularity, just as in the case
of electrodynamics where the divergence occurred in the classical
Coulomb potential gets resolved in the realm of quantum electrodynamics.

However in the absence of a fully accepted quantum gravity, one may
think of some classical resolution of the disturbing aspects of such
singularities. One motivation in this regard comes through the Born-Infeld
(BI) theory of classical electrodynamics \citep{Born:1934gh}, where
the self-energy of the electron, as well as the field strength, proves
to be bounded, and hence remove the divergences occurred in the standard
Maxwell theory. The gravitational extension of Born-Infeld theory
has been explored in \citep{Deser:1998rj,Banados:2010ix} and shows
some interesting consequences in cosmology as well as in black hole
physics. It turns out that the Born-Infeld gravitational theory can
trigger a non-singular cosmic bounce in simple scenarios involving
a radiation fluid or a pressureless dust fluid, which in turn makes
the universe's evolution free from any curvature singularity \citep{Banados:2010ix}.
Consequently, the implications of Born-Infeld gravitational theory
were extensively examined in cosmology \citep{Odintsov:2014yaa,Du:2014jka,Kim:2013noa,Kruglov:2013qaa,Yang:2013hsa,Avelino:2012ue,Escamilla-Rivera:2012jfb,Cho:2012vg,Scargill:2012kg},
astrophysics \citep{Harko:2013xma,Avelino:2012ge}, black holes \citep{Olmo:2013gqa}
and wormhole physics \citep{Lobo:2014fma,Harko:2013aya}. Attempts
have been made to generalize the BI gravitational theory in the context
of a higher curvature scenario, by adding a $f(R)$ Lagrangian to
the original theory \citep{Makarenko:2014lxa,Makarenko:2014nca} (one
may go through \citep{Nojiri:2010wj,Nojiri:2017ncd,Capozziello:2011et,Nojiri:2013zza,delaCruz-Dombriz:2012bni}
for the reviews on standard $f(R)$ gravity). The $f(R)$ piece in
the BI-$f(R)$ gravity acts as some new curvature interactions which
has interesting implications. For instance, it was found that the
cosmic bouncing solutions are robust against the BI-$R^{2}$ gravity
theory, i.e., when the quadratic $f(R)$ serves the higher curvature
part in the BI-$f(R)$ theory.

On other hand, unimodular Einstein-Hilbert gravity \citep{Finkelstein:2000pg,Alvarez:2005iy,Abbassi:2007bq,Ellis:2010uc,Jain:2012cw,SINGH_2013,Kluson:2014esa,Padilla:2014yea,Barcelo:2014mua,Barcelo:2014qva,burger2015klt,Alvarez:2015sba,Jain:2012gc,Alvarez:2023utn}
carries a lot of success as cosmological constant appears in such
theory without adding it by hand, unlike to the standard Einstein-Hilbert
theory where the cosmological constant is put by hand. In particular,
the cosmological constant in the unimodular theory originates from
the trace-free part of the Einstein field equations. In the context
of unimodular gravity, the determinant of spacetime metric is constrained
to be unity, which is suitably adjusted by incorporating a Lagrange
multiplier in the gravitational action. Regarding the cosmological
implications, the unimodular gravity serves as a good avenue for late-time
acceleration where the spacetime metric is decomposed into two parts,
the unimodular metric and a scalar field \citep{Jain:2012gc,Jain:2011jc}.
Owing to such interesting implications, the unimodular theory is extended
in the context of $f(R)$ gravity --- dubbed as unimodular $f(R)$
gravity \citep{Nojiri:2015sfd,Nojiri:2016plt} where the basic assumption
of unimodular theory, namely the determinant of the metric is unity,
is used by introducing a Lagrange multiplier to the $f(R)$ part of
the action. The unimodular $f(R)$ gravity proves a successful stand
in the cosmology sector, see \citep{Nojiri:2015sfd,Nojiri:2016ygo}.

Having the two different theories in hand, namely the BI-$f(R)$ theory
and the unimodular $f(R)$ theory, and owing to their qualitatively
interesting cosmological implications, the present work is devoted
to exploring the unimodular generalization of the Born-Infeld-$f(R)$
gravity within the framework of cosmology. After formulating the action
corresponding to the unimodular Born-Infeld-$f(R)$ gravity, we will
present a reconstruction scheme of this unimodular extension to achieve
various cosmological eras of the universe. Interestingly, the unimodular
BI-$f(R)$ theory serves as a good agent for various cosmic evolutions
of the universe (including the early phase of the universe), which
are also in agreement with the recent Planck data. The potential applications
of the reconstruction method we will present in this paper are quite
many since it is possible to realize various cosmological evolutions,
which are exotic for the standard Einstein-Hilbert gravity and cannot
be realized in that case, for example bouncing cosmologies.

The paper is organized as follows: after a brief discussion of the
Born-Infeld-$f(R)$ theory in Sec.~{[}\ref{sec-2}{]}, we will formulate
the unimodular generalization of the Born-Infeld-$f(R)$ theory and
its various implications in the sections afterwards. For this purpose,
we will investigate which metric proves to be compatible with the
unimodular constraint since the standard FRW metric is unable to do
so. The paper ends with some conclusions and future proposals in Sec.~{[}\ref{sec-conclusion}{]}.

\section{Born-Infeld-$f\left(R\right)$ theory}

\label{sec-2}

Let us briefly review the standard BI-$f\left(R\right)$ theory \citep{Makarenko:2014lxa}
(see also \citep{Makarenko:2014nca,Odintsov:2014yaa}). In this theory,
the original Einstein-BI gravity theory is combined with an additional
$f\left(R\right)$ that depends on the Ricci scalar $R=g^{\mu\nu}R_{\mu\nu}\left(\Gamma\right)$.
To avoid any ghost instabilities, the theory is formulated within
the Palatini formalism (for more detail see \citep{BeltranJimenez:2017doy}),
in which the metric $g_{\mu\nu}$ and the connection $\Gamma_{\beta\gamma}^{\alpha}$
are treated as independent variables. The action for this theory is
given by

\begin{eqnarray}
S & = & \frac{2}{\kappa^{2}\epsilon}\int d^{4}x\left[\sqrt{-|g_{\mu\nu}+\epsilon R_{\mu\nu}|}-\lambda\sqrt{-g}\right]+\frac{\alpha}{2\kappa^{2}}\int d^{4}x\left[\sqrt{-g}f\left(R\right)\right]+S_{m},\label{eq: action_f(R)}
\end{eqnarray}
where the first term represents the standard BI gravitational Lagrangian
and the second term is an additional function of the Ricci scalar.
$S_{m}$ is the matter action which depends on generically $\psi$
field and the metric tensor $g_{\mu\nu}$.

\begin{equation}
R_{\mu\beta\nu}^{\alpha}=\partial_{\beta}\Gamma_{\nu\mu}^{\alpha}-\partial_{\nu}\Gamma_{\mu\beta}^{\alpha}+\Gamma_{\beta\lambda}^{\alpha}\Gamma_{\nu\mu}^{\lambda}-\Gamma_{\nu\lambda}^{\alpha}\Gamma_{\mu\beta}^{\lambda},
\end{equation}
is the Riemann tensor of the connection $\Gamma_{\mu\nu}^{\lambda}$.
In addition, the connection is also assumed to be torsionless. The
parameter $\kappa$ is a constant with inverse dimensions to that
of a cosmological constant and $\lambda$ is a dimensionless constant.
Note that $g$ is the determinant of the metric. Throughout this paper,
we will use Planck units $8\pi G=1$ and set the speed of light to
$c=1$.

The variation of this action with respect to the metric tensor leads
to a modified metric field equations for the standard BI gravitational
model, 
\begin{equation}
\frac{\sqrt{-q}}{\sqrt{-g}}\left(q^{-1}\right)^{\mu\nu}-g^{\mu\nu}\left(\lambda-\frac{\alpha\epsilon}{2}f\left(R\right)\right)-\alpha\epsilon f_{R}R^{\mu\nu}=-\kappa^{2}\epsilon T^{\mu\nu},\label{eq: eom_g-1}
\end{equation}
where $f_{R}$ is the derivative of $f\left(R\right)$ with respect
to the Ricci scalar, and $T^{\mu\nu}$ is the standard energy-momentum
tensor. We have used the notation, 
\begin{equation}
q_{\mu\nu}=g_{\mu\nu}+\epsilon R_{\mu\nu}.\label{eq: q_f(R)}
\end{equation}
We denoted the inverse of $q_{\mu\nu}$ by $\left(q^{-1}\right)^{\mu\nu}$
and $q$ represents the determinant of $q_{\mu\nu}$. Similarly, the
corresponding equation which follows by variation over the connection
$\Gamma$ has the form,

\begin{equation}
\nabla_{\lambda}\left(\sqrt{-q}q^{\mu\nu}+\alpha\sqrt{-g}f_{R}g^{\mu\nu}\right)=0,\label{eq: eom_conn}
\end{equation}
where the covariant derivative is taken with respect to the connection
which is defined for a scalar field $\phi$ as 
\begin{equation}
\nabla_{\mu}\phi=\partial_{\mu}\phi-\Gamma_{\mu\alpha}^{\alpha}\phi.
\end{equation}

\section{Unimodular generalization}

Let us now present briefly the basic idea of unimodular $f\left(R\right)$
gravity \citep{Nojiri:2015sfd}. This approach is based on the assumption
that the metric satisfies the following constraint

\begin{equation}
\sqrt{-g}=1.\label{eq: UnimodularCons}
\end{equation}
In addition, we assume that the metric expressed in terms of the cosmological
time $t$ is a flat Friedman-Robertson-Walker (FRW) of the form,

\begin{eqnarray}
ds^{2} & = & -dt^{2}+a\left(t\right)^{2}\left(dx^{2}+dy^{2}+dz^{2}\right).\label{eq: FLRW metric}
\end{eqnarray}
The metric (\ref{eq: FLRW metric}) does not satisfy the unimodular
constraint (\ref{eq: UnimodularCons}), we redefined the cosmological
time $t$, to a new variable $\tau$, as follows: 
\begin{equation}
d\tau=a\left(t\right)^{3}dt,\label{eq:d_tau}
\end{equation}
in which case, the metric of Eq.(\ref{eq: FLRW metric}), becomes
the ``unimodular metric'', 
\begin{equation}
ds^{2}=-a\left(t\left(\tau\right)\right)^{-6}d\tau^{2}+a\left(t\left(\tau\right)\right)^{2}\left(dx^{2}+dy^{2}+dz^{2}\right).\label{eq: UnimodularMetric}
\end{equation}
and hence the unimodular constraint is satisfied. Note that we use
$a\left(\tau\right)=a\left(t\left(\tau\right)\right)$. For the unimodular
metric of Eq.(\ref{eq: UnimodularMetric}), the non-vanishing components
of the Levi-Civita connection and of the curvatures are given below,
\begin{equation}
\Gamma_{tt}^{t}=-3K,\,\,\,\,\,\,\,\,\,\,\Gamma_{ij}^{t}=a^{8}K\delta_{ij},\,\,\,\,\,\,\,\,\,\,\,\Gamma_{jt}^{i}=\Gamma_{tj}^{i}=K\delta_{j}^{i},
\end{equation}

\begin{equation}
R_{tt}=-3\dot{K}-12K^{2},\,\,\,\,\,\,\,\,\,\,R_{ij}=a^{8}\left(\dot{K}+6K^{2}\right)\delta_{ij},
\end{equation}
where the ``dot'' denoting as usual differentiation with respect
to $\tau$ and we introduced the function $K\left(\tau\right)$, which
is equal to $K=\frac{1}{a}\frac{da}{d\tau}$ , so it is the analog
Hubble rate for the $\tau$ variable. In addition, the corresponding
Ricci scalar curvature corresponding to the metric (\ref{eq: UnimodularMetric}),
is equal to,

\begin{equation}
R=a^{6}\left(6\dot{K}+30K^{2}\right).
\end{equation}
In the following, we shall refer to the metric of Eq.(\ref{eq: UnimodularMetric})
as the unimodular FRW metric. By making use of the Lagrange multiplier
method, the vacuum Jordan frame unimodular $f\left(R\right)$ gravity
action is given by 
\begin{equation}
S=\int d^{4}x\left\{ \left[\sqrt{-g}\left(f\left(R\right)-\gamma\right)\right]+\gamma\right\} ,\label{eq: Unimodular_FR}
\end{equation}
where $\gamma\left(x\right)$ stands for the Lagrange multiplier function.
If one take the variation with respect to $\gamma\left(x\right)$,
it can be seen the unimodular constraint in Eq.(\ref{eq: UnimodularCons})
is satisfied.

From this idea, we modify the action in Eq.(\ref{eq: action_f(R)})

\begin{eqnarray}
S & = & \frac{2}{\kappa^{2}\epsilon}\int d^{4}x\left[\sqrt{-|g_{\mu\nu}+\epsilon R_{\mu\nu}|}-\lambda\sqrt{-g}\right]+\frac{\alpha}{2\kappa^{2}}\int d^{4}x\left\{ \left[\sqrt{-g}\left(f\left(R\right)-\gamma\right)\right]+\gamma\right\} +S_{m},\label{eq: action_f(R)-1}
\end{eqnarray}
The variation of this action with respect to the metric tensor, 
\begin{equation}
\frac{\sqrt{-q}}{\sqrt{-g}}q^{\mu\nu}-\left[\lambda-\frac{\alpha\epsilon}{2}\left(f\left(R\right)-\gamma\right)\right]g^{\mu\nu}-\alpha\epsilon f_{R}R^{\mu\nu}=-\kappa^{2}\epsilon T^{\mu\nu},\label{eq: eom_g}
\end{equation}
Similarly, $q_{\mu\nu}$ has the same definition with Eq.(\ref{eq: q_f(R)})
and the corresponding equation which follows by variation over the
connection $\Gamma$ has the same form with Eq.(\ref{eq: eom_conn})
where the covariant derivative is defined in terms of the independent
connection $\Gamma_{\beta\gamma}^{\alpha}$. 

\subsection{Conformal approach and applications of the reconstruction method}

If we assume that $q_{\mu\nu}$ is conformally proportional to the
metric tensor as 
\begin{equation}
q_{\mu\nu}=k\left(\tau\right)g_{\mu\nu},\label{eq: conf_rel}
\end{equation}
then Eq.(\ref{eq: eom_conn}) becomes, 
\begin{equation}
\nabla_{\nu}\left[\sqrt{-u}u^{\mu\nu}\right]=0.\label{eq: eom_u}
\end{equation}
In this case, we have an auxiliary metric defined as $u_{\mu\nu}=\left[k\left(\tau\right)+f_{R}\right]g_{\mu\nu}$
and $u^{\mu\nu}$ represents the inverse representation of $u_{\mu\nu}$.
Eq.(\ref{eq: eom_u}) tells us that the connection can be defined
by this auxiliary metric as 
\begin{equation}
\Gamma_{\mu\nu}^{\rho}=\frac{1}{2}u^{\rho\sigma}\left(u_{\sigma\nu,\mu}+u_{\mu\sigma,\nu}-u_{\mu\nu,\sigma}\right).
\end{equation}
By considering the conformal relation between $q_{\mu\nu}$ and $g_{\mu\nu}$
in Eq.(\ref{eq: conf_rel}), we can easily say that the Ricci tensor
must also be proportional to the metric tensor. Thus one can write
the relationship between the Ricci tensor and $g_{\mu\nu}$, as 
\begin{equation}
R_{\mu\nu}=\frac{1}{\kappa}\left[k\left(\tau\right)-1\right]g_{\mu\nu}.\label{eq: Ricci_g}
\end{equation}
For a cosmological scenario, we use the unimodular metric in Eq.(\ref{eq: UnimodularMetric}).
Now, we can define the auxiliary metric as 
\begin{equation}
u_{\mu\nu}=u\left(\tau\right)\text{diag}\left(-a\left(\tau\right)^{-6},a^{2}\left(\tau\right),a^{2}\left(\tau\right),a^{2}\left(\tau\right)\right),\label{eq: metric_U}
\end{equation}
where 
\begin{eqnarray}
u\left(\tau\right) & = & k\left(\tau\right)+f_{R}.
\end{eqnarray}
According to Eq.(\ref{eq: Ricci_g}), we can define the Ricci tensor
as $R_{\mu\nu}=r\left(\tau\right)g_{\mu\nu}$. From this background,
one can find the following expressions 
\begin{equation}
r\left(\tau\right)=\frac{3}{2}a^{6}\left[2\frac{\ddot{a}}{a}+4\frac{\dot{a}}{a}\frac{\dot{u}}{u}+\frac{\ddot{u}}{u}-\left(\frac{\dot{u}}{u}\right)^{2}+6\left(\frac{\dot{a}}{a}\right)^{2}\right],
\end{equation}

\begin{equation}
r\left(\tau\right)=a^{6}\left[\frac{\ddot{a}}{a}+4\frac{\dot{a}}{a}\frac{\dot{u}}{u}+\frac{\ddot{u}}{2u}+5\left(\frac{\dot{a}}{a}\right)^{2}\right].
\end{equation}
Then using these two equations, one can find the following expressions
as well, 
\begin{equation}
r\left(\tau\right)=3a^{6}\left(\frac{\dot{a}}{a}+\frac{\dot{u}}{2u}\right)^{2},\label{eq: r(tau)_3}
\end{equation}
\begin{equation}
2\frac{\ddot{a}}{a}+4\left(\frac{\dot{a}}{a}\right)^{2}+2\frac{\dot{a}}{a}\frac{\dot{u}}{u}-\frac{3}{2}\left(\frac{\dot{u}}{u}\right)^{2}+\frac{\ddot{u}}{u}=0.\label{eq: u_a_equ}
\end{equation}
The last equation shows an exact relation between $a\left(\tau\right)$
and $u\left(\tau\right)$. If we solve Eq.(\ref{eq: u_a_equ}) with
respect to $u\left(\tau\right)$, we get the following relation 
\begin{equation}
u\left(\tau\right)=\frac{4}{\mathcal{W}^{2}},\label{eq: u_val}
\end{equation}
where 
\begin{equation}
\mathcal{W}\left(\tau\right)=a\left(\tau\right)\left(C_{1}\int\frac{1}{a\left(\tau\right)^{4}}d\tau-C_{2}\right),
\end{equation}
where $C_{1}$ and $C_{2}$ are integration constants. Note that constants
$C_{1}$ and $C_{2}$ do not have real physical meaning in this equation.
However, later on they enter to final form of $f(R)$ and hence their
role is in the form of $f(R)$ which corresponds to given cosmology.
Then by using this solution, Eq.(\ref{eq: r(tau)_3}) becomes 
\begin{equation}
r\left(\tau\right)=\frac{3C_{1}^{2}}{\mathcal{W}^{2}}.\label{eq: r(tau)}
\end{equation}
Now, from the previous definitions, it can also be shown that 
\begin{equation}
r\left(\tau\right)=\frac{1}{\kappa}\left(u\left(\tau\right)-f_{R}-1\right),\label{eq: r(tau)_def}
\end{equation}
so equating this expression with Eq.(\ref{eq: r(tau)}) we get, 
\begin{equation}
f_{R}=\frac{4-3\kappa C_{1}^{2}}{\mathcal{W}^{2}}-1,\label{eq: fR_val}
\end{equation}
then the general definition of $f\left(R\right)$ function can be
found as follows 
\begin{equation}
f\left(R\right)=\left(\frac{4-3\kappa C_{1}^{2}}{\mathcal{W}^{2}}-1\right)R+C_{3},\label{eq: f_val}
\end{equation}
here $C_{3}$ is another integration constant. Using Eq.(\ref{eq: eom_g})
one can easily find the unimodular Lagrange multiplier $\gamma\left(\tau\right)$
is written 
\begin{equation}
\gamma\left(\tau\right)=f\left(R\right)-2\left(\frac{1}{\alpha\epsilon}+\frac{1}{r\left(\tau\right)}\right)f_{R}+\frac{2}{\alpha\epsilon}\left(u\left(\tau\right)-\lambda\right),\label{eq: gamma(tau)}
\end{equation}
Finally, using Eqs.(\ref{eq: u_val}), (\ref{eq: r(tau)}), (\ref{eq: fR_val})
and (\ref{eq: f_val}), it can also written as below, 
\begin{equation}
\gamma\left(\tau\right)=\left(\frac{4-3\kappa C_{1}^{2}}{\mathcal{W}^{2}}-1\right)\left(\frac{2R-2}{\alpha\epsilon}-\frac{2\mathcal{W}^{2}}{3C_{1}^{2}}\right)+\frac{8}{\alpha\epsilon\mathcal{W}^{2}}-\frac{2}{\alpha\epsilon}\lambda.
\end{equation}
Having described all the essential features of the unimodular Born-Infeld-$f\left(R\right)$
gravity and the corresponding reconstruction method, let us now demonstrate
how various cosmological scenarios can be realized in the context
of unimodular Born-Infeld-$f\left(R\right)$ gravity.

\subsubsection{de Sitter expanding Universe}

Let us for example consider the case that the FRW metric describes
a de Sitter expanding Universe, for which the scale factor is, 
\begin{equation}
a\left(t\right)=e^{H_{0}t},
\end{equation}
where $H_{0}$ is the Hubble parameter and is a constant. In this
case, by using Eq.(\ref{eq:d_tau}), we get the relations between
$t$ and $\tau$ as follows 
\begin{equation}
\tau\left(t\right)=\frac{e^{3H_{0}t}}{3H_{0}},\,\,\,\,\,\,t\left(\tau\right)=\frac{\ln\left(3H_{0}\tau\right)}{3H_{0}},
\end{equation}
then the scale factor as a function of unimodular time $\tau$ 
\begin{equation}
a\left(\tau\right)=\left(3H_{0}\tau\right)^{\frac{1}{3}}.
\end{equation}
Using this expression, one can find the following equation by solving
Eq.(\ref{eq: u_a_equ})

\begin{equation}
u\left(\tau\right)=\frac{4}{\left(3C_{1}\tau^{1/3}+C_{2}\right)^{2}},
\end{equation}
then taking account of this solution, Eq.(\ref{eq: r(tau)_3}) becomes

\begin{equation}
r\left(\tau\right)=\frac{3H_{0}^{2}C_{2}^{2}}{\left(3C_{1}\tau^{1/3}+C_{2}\right)^{2}}.
\end{equation}
Now, equating Eq.(\ref{eq: r(tau)_def}) with Eq.(\ref{eq: r(tau)})
we get, 
\begin{equation}
f_{R}=\frac{-3\kappa H_{0}^{2}C_{2}^{2}+4}{\left(3C_{1}\tau^{1/3}+C_{2}\right)^{2}}-1,
\end{equation}
then the $f\left(R\right)$ function can be found as follows 
\begin{equation}
f\left(R\right)=\left[\frac{-3\kappa H_{0}^{2}C_{2}^{2}+4}{\left(3C_{1}\tau^{1/3}+C_{2}\right)^{2}}-1\right]R+C_{3},
\end{equation}
where $C_{3}$ is an integration constant. From this point, by using
Eq.(\ref{eq: gamma(tau)}) one can easily find the unimodular Lagrange
multiplier $\gamma\left(\tau\right)$ is

\begin{equation}
\gamma\left(\tau\right)=\frac{\left(18-9\alpha\epsilon R\right)\kappa\sigma^{2}+\left\{ -3\alpha\epsilon\left[\left(R-2\kappa-C_{3}\right)\mathcal{W}-4R\right]-6\mathcal{W}\left(\lambda-1\right)\right\} \sigma+2\alpha\epsilon\left(\mathcal{W}-4\right)\mathcal{W}}{3\alpha\epsilon\sigma\mathcal{W}},
\end{equation}
where $\mathcal{W}\left(\tau\right)$ and $\sigma$ are defined as
follows
\begin{equation}
\mathcal{W}\left(\tau\right)=\left(3C_{1}\tau^{1/3}+C_{2}\right)^{2},
\end{equation}
\begin{equation}
\sigma=H_{0}^{2}C_{2}^{2}.
\end{equation}

Therefore the Unimodular Born-Infeld-$f\left(R\right)$ gravity can
lead to a de-Sitter inflationary scenario during the early universe,
provided the $f(R)$ and the $\gamma(\tau)$ have the above forms.
However, a de-Sitter inflation has no exit mechanism, i.e., the inflation
becomes eternal; and moreover the scalar spectral index of primordial
curvature perturbation gets exactly scale invariant in the context
of a de-Sitter inflation, which is not consistent with the recent
Planck data at all. This indicates that the constant Hubble parameter
corresponding to the de-Sitter case does not lead to a good inflationary
scenario of the universe. 

On the other hand, eternal inflation, as demonstrated in the context
of de-Sitter inflation, carries significant implications for our understanding
of the cosmic landscape. This perpetual inflationary phase suggests
the potential existence an expansive multiverse with infinite pocket
universes \citep{Guth:2007EternalInflation}, enriching our understanding
of cosmic origins and structure within the broader cosmological framework.

\subsubsection{The Starobinsky inflation model}

In order to achieve a viable inflation, let us consider the Starobinsky
inflation with relatively interesting inflationary phenomenology,
where the Hubble rate as a function of the cosmic time is given by,

\begin{equation}
H\left(t\right)=H_{0}-\frac{M^{2}}{6}\left(t-t_{i}\right),\label{H1}
\end{equation}
with $H_{0}$, $M$, and $t_{i}$ are arbitrary constants. Actually
$t_{i}$ represents the beginning of inflation, which can be considered
as the horizon crossing of the large scale mode ($\sim0.05\mathrm{Mpc}^{-1}$),
and moreover, $H_{0}$ sets the inflationary energy scale at its onset.
The slow roll parameter, defined by $\epsilon_{\mathrm{1}}=-\dot{H}/H^{2}$,
corresponding to the above Hubble parameter comes as, 
\begin{equation}
\epsilon_{\mathrm{1}}=\frac{M^{2}}{6\left\{ H_{0}^{2}-\frac{M^{2}}{3}\left(t-t_{i}\right)\right\} }.\label{e1}
\end{equation}
Clearly, for $M<H_{0}$, the slow roll parameter is less than unity
at $t=t_{i}$ which indicates an accelerated phase of the universe.
Moreover, Eq.~(\ref{e1}) also indicates that $\epsilon_{\mathrm{1}}$
is an increasing function of time, and eventually reaches to unity
at 
\begin{equation}
t_{\mathrm{f}}=t_{i}+\frac{3H_{0}}{M^{2}}\left(H_{0}^{2}-\frac{M^{2}}{6}\right),\label{exit time}
\end{equation}
which describes the end of inflation. Thereby unlike to the de-Sitter
case, the Starobinsky inflation (where the Hubble parameter follows
Eq.~(\ref{H1})) provides an exit of inflation. Furthermore the inflationary
indices in the Starobinsky inflation, particularly the spectral index
of scalar perturbation ($n_{s}$) and the tensor-to-scalar ratio ($r$)
have the well known expressions: 
\[
n_{s}=1-\frac{2}{N}~~~~~~~\mathrm{and}~~~~~~~r=\frac{12}{N^{2}},
\]
where $N$ is the total e-folding number of inflationary period. For
$N\approx60$, such observable indices are in good agreement with
the Planck data \citep{Planck:2018jri}, which makes the Starobinsky
inflation a viable one.

In the realm of normal $f(R)$ theory, this model corresponds to the
well known $R^{2}$ inflation Starobinsky model \citep{Starobinsky:1980ANew,Barrow:1988Inflation,Odintsov:2015BouncingCosmology,Odintsov:2015SingularInflationary},
and the corresponding scale factor is, 
\begin{equation}
a\left(t\right)=a_{0}e^{H_{0}\left(t-t_{i}\right)-\frac{M^{2}}{12}\left(t-t_{i}\right)}.
\end{equation}
Since the study of this model is done for early times, we can approximate
this scale factor as follows,

\begin{equation}
a\left(t\right)=a_{0}e^{H_{0}\left(t-t_{i}\right)+\frac{M^{2}}{6}tt_{i}},
\end{equation}
which easily follows since $t,t_{i}\ll1/H_{0}$. Then, by using Eq.(\ref{eq:d_tau}),
we easily obtain that the coordinate $\tau$ is related to the cosmic
time $t$, as follows, 
\begin{equation}
\tau\left(t\right)=\frac{2a_{0}^{3}e^{-3H_{0}t_{i}+\frac{1}{2}t\left(6H_{0}+M^{2}t_{i}\right)}}{6H_{0}+M^{2}t_{i}},
\end{equation}
which can be easily inverted to yield the function $t=t\left(\tau\right)$,
which is,

\begin{equation}
t\left(\tau\right)=\frac{2\left(3H_{0}t_{i}+\ln\left(\frac{\left(6H_{0}+M^{2}t_{i}\right)\left(\tau-\tau_{0}\right)}{2a_{0}^{3}}\right)\right)}{6H_{0}+M^{2}t_{i}},\label{eq:t_tau}
\end{equation}
with $\tau_{0}$ being a fiducial initial time instance. Having Eq.(\ref{eq:t_tau}),
we can express the scale factor and the Hubble rate as functions of
$\tau$, so the scale factor $a\left(\tau\right)$ is equal to, 
\begin{equation}
a\left(\tau\right)=\left(\frac{\left(6H_{0}+M^{2}t_{i}\right)\left(\tau-\tau_{0}\right)}{2}\right)^{\frac{1}{3}},\label{eq: a_tau}
\end{equation}
and solving Eq.(\ref{eq: u_a_equ}) we obtain 
\begin{equation}
u\left(\tau\right)=\frac{4}{\left(3C_{1}\left(\tau-\tau_{0}\right)^{1/3}+C_{2}\right)^{2}},
\end{equation}
Using these metric expressions Eq.(\ref{eq: r(tau)_3}) takes the
following form

\begin{equation}
r\left(\tau\right)=\frac{C_{2}^{2}\left(6H_{0}+M^{2}t_{i}\right)^{2}}{12\left(3C_{1}\left(\tau-\tau_{0}\right)^{1/3}+C_{2}\right)^{2}}.
\end{equation}
Now, we can find the $f_{R}$ expression by using Eqs.(\ref{eq: r(tau)_def})
and (\ref{eq: r(tau)}), 
\begin{equation}
f_{R}=\frac{48-\kappa C_{2}^{2}\left(6H_{0}+M^{2}t_{i}\right)^{2}}{12\left(3C_{1}\left(\tau-\tau_{0}\right)^{1/3}+C_{2}\right)^{2}}-1,
\end{equation}
then solving this equation with respect to $f\left(R\right)$ function,
we can find the exact formulation as,
\begin{equation}
f\left(R\right)=\left[\frac{48-\kappa C_{2}^{2}\left(6H_{0}+M^{2}t_{i}\right)^{2}}{12\left(3C_{1}\left(\tau-\tau_{0}\right)^{1/3}+C_{2}\right)^{2}}-1\right]R+C_{3},\label{f1}
\end{equation}
where $C_{3}$ is an integration constant. From this point, by using
Eq.(\ref{eq: gamma(tau)}) one can easily find the unimodular Lagrange
multiplier $\gamma\left(\tau\right)$ is

\begin{equation}
\gamma\left(\tau\right)=\frac{\kappa C_{2}^{4}\sigma^{2}\left(2-\alpha\epsilon R\right)-12C_{2}^{2}\sigma\left\{ \alpha\epsilon\left[\left(R-C_{3}-\frac{\kappa}{6}\right)\mathcal{W}-\frac{13R}{3}\right]+\frac{2}{3}+2\left(\lambda-1\right)\mathcal{W}\right\} +24\alpha\epsilon\left(\mathcal{W}-\frac{13}{3}\right)\mathcal{W}}{12\alpha\epsilon C_{2}^{2}\sigma\mathcal{W}},\label{gamma1}
\end{equation}
where 

\begin{equation}
\mathcal{W}\left(\tau\right)=\left(3\left(\tau-\tau_{0}\right)^{1/3}C_{1}+C_{2}\right)^{2},
\end{equation}
\begin{equation}
\sigma=\left(6H_{0}+M^{2}t_{i}\right)^{2}.
\end{equation}

Therefore the $f(R)$ of Eq.(\ref{f1}) and the $\gamma(\tau)$ of
Eq.(\ref{gamma1}) result to a Starobinsky-type inflation (where the
Hubble parameter takes the form as of Eq.(\ref{H1})) in the present
context of Unimodular Born-Infeld-$f\left(R\right)$ gravity.

\subsubsection{Power-law expansion factor}

Consider another example, for which the Universe is described by a
power-law scale factor of the form,

\begin{equation}
a\left(t\right)=\left(\frac{t}{t_{0}}\right)^{f_{0}},\label{eq: a_3}
\end{equation}
where $t_{0}$ and $f_{0}$ are constants. The constant $f_{0}$ sets
the effective EoS parameter ($w$) corresponding to the above scale
factor by $f_{0}=\frac{2}{3\left(1+w\right)}$. The power law scale
factor is interesting to study, as it describes the universe's evolution
after the end of inflation. In particular, $f_{0}=\frac{2}{3}$ represents
a matter dominated universe, while radiation dominated era can be
described by $f_{0}=\frac{1}{2}$. Apart from these two standard era,
the perturbative reheating era (that connects the inflation with the
Standard Big-Bang cosmology) may also be described by such power law
scale factor. The effective EoS parameter during the reheating era
(or equivalently, the $f_{0}$) can take $0<w<1$ depending on the
dynamics.\\
 For the power law scale factor, by using Eq.(\ref{eq:d_tau}), we
get, 
\begin{equation}
\tau\left(t\right)=\frac{t_{0}}{3f_{0}+1}\left(\frac{t}{t_{0}}\right)^{3f_{0}+1},\,\,\,\,\,\,\,\,\,t\left(\tau\right)=t_{0}\left(\frac{\left(3f_{0}+1\right)}{t_{0}}\tau\right)^{\frac{1}{3f_{0}+1}},
\end{equation}
and by substituting in Eq.(\ref{eq: a_3}), we easily get the scale
factor in terms of the variable $\tau$, 
\begin{equation}
a\left(\tau\right)=\left(\frac{\tau}{\tau_{0}}\right)^{h_{0}},\label{eq: a(tau)_power}
\end{equation}
where $h_{0}=\frac{f_{0}}{3f_{0}+1}$ and $\tau_{0}=\frac{t_{0}}{3f_{0}+1}$.
By using this relation, solving Eq.(\ref{eq: u_a_equ}) we get

\begin{equation}
u\left(\tau\right)=\frac{4\tau^{6h_{0}}\left(4h_{0}-1\right)^{2}}{\left(C_{1}\tau^{4h_{0}}+C_{2}\tau\right)^{2}},
\end{equation}
then substituting this result in Eq.(\ref{eq: r(tau)_3}), we can
find

\begin{equation}
r\left(\tau\right)=\frac{3C_{2}^{2}\left(\frac{\tau}{\tau_{0}}\right)^{6h_{0}}\left(4h_{0}-1\right)^{2}}{\left(C_{1}\tau^{4h_{0}}+C_{2}\tau\right)^{2}}.
\end{equation}
Now, similar to the previous calculations, if we combine Eq.(\ref{eq: r(tau)_def})
with Eq.(\ref{eq: r(tau)}) we get, 
\begin{equation}
f_{R}=\frac{\left(4h_{0}-1\right)^{2}\left(-3C_{2}^{2}\left(\frac{\tau}{\tau_{0}}\right)^{6h_{0}}+4\tau^{6h_{0}}\right)}{\left(C_{1}\tau^{4h_{0}}+C_{2}\tau\right)^{2}}-1,
\end{equation}
then the $f\left(R\right)$ function can also be found as 
\begin{equation}
f\left(R\right)=-\left[\frac{\left(4h_{0}-1\right)^{2}\left(-3C_{2}^{2}\left(\frac{\tau}{\tau_{0}}\right)^{6h_{0}}+4\tau^{6h_{0}}\right)}{\left(C_{1}\tau^{4h_{0}}+C_{2}\tau\right)^{2}}-1\right]R+C_{3},
\end{equation}
where $C_{3}$ is an integration constant. The unimodular Lagrange
multiplier $\gamma\left(\tau\right)$ can be found by the help of
Eq.(\ref{eq: gamma(tau)}),

\begin{equation}
\gamma\left(\tau\right)=\frac{2\left(\mathcal{W}-4\sigma\tau^{6h_{0}}\right)\left(\frac{\tau}{\tau_{0}}\right)^{-6h_{0}}}{3C_{2}^{2}\sigma}+\frac{4\alpha\epsilon\sigma\tau^{6h_{0}}R-3\kappa C_{2}^{2}\sigma\left(\alpha\epsilon R-2\right)\left(\frac{\tau}{\tau_{0}}\right)^{6h_{0}}-\mathcal{W}\left[\alpha\epsilon\left(R-2\kappa-C_{3}\right)+2\lambda-2\right]}{\alpha\epsilon\mathcal{W}},
\end{equation}
where 

\begin{equation}
\mathcal{W}\left(\tau\right)=\left(C_{1}\tau^{4h_{0}}+C_{2}\tau\right)^{2},
\end{equation}
\begin{equation}
\sigma=\left(4h_{0}-1\right)^{2}.
\end{equation}

The above three subsections describe the required form of $f(R)$
and the unimodular Lagrange multiplier to achieve inflation and the
Standard Big Bang evolution of the universe. However, we would like
to mention that we have not described the universe's evolution as
a unified picture. It will be interesting to address the unification
of various cosmological eras in the context of unimodular $f(R)$
gravity, which we expect to study elsewhere.

\subsubsection{Bounce scenario}

Another scenario with interesting phenomenology is the superbounce
scenario \citep{Koehn:2014Cosmological,Odintsov:2015Superbounce,Oikonomou:2015Superbounce,Odintsov:2022unp},
which was firstly studied in the context of some ekpyrotic scenarios
\citep{Koehn:2014Cosmological}. The scale factor and the Hubble rate
for the superbounce are given below,

\begin{equation}
a\left(t\right)=\left(-t+t_{s}\right)^{\frac{2}{c^{2}}},\,\,\,\,\,\,H\left(t\right)=-\frac{2}{c^{2}\left(-t+t_{s}\right)},\label{eq: a_3-1}
\end{equation}
with $c$ being an arbitrary parameter of the theory while the bounce
in this case occurs at $t=t_{s}$. In this case, by using Eq.(\ref{eq:d_tau}),
we get, 
\begin{equation}
\tau\left(t\right)=\frac{c^{2}\left(-t+t_{s}\right)^{\frac{\left(c^{2}+6\right)}{c^{2}}}}{c^{2}+6},\,\,\,\,\,\,\,\,\,t\left(\tau\right)=-\left(-\frac{c^{2}}{\left(c^{2}+6\right)\tau}\right)^{\frac{2}{c^{2}+6}}+t_{s},
\end{equation}
and by substituting in Eq.(\ref{eq: a_3-1}), we easily get the scale
factor in terms of the variable $\tau$,

\begin{equation}
a\left(\tau\right)=\left(\frac{\tau}{\tau_{0}}\right)^{h_{0}},
\end{equation}
where $h_{0}=\frac{2}{c^{2}+6}$, $\tau_{0}=\frac{c^{2}}{c^{2}+6}$.
This is the same structure with the power law expansion case in Eq.(\ref{eq: a(tau)_power}).
So we can say that the power law expansion of the scale factor in
unimodular time lead to a bounce scenario for the evolution of our
universe.

\subsubsection{Exponential evolution of $u\left(\tau\right)$}

Let us consider, the auxiliary metric function evolve as $u\left(\tau\right)=e^{h\tau}$.
In this case, If we solve Eq.(\ref{eq: u_a_equ}) with respect to
the $a\left(\tau\right)$ then the scale factor becomes 
\begin{equation}
a\left(\tau\right)=\left[\frac{3}{2h}\left(C_{1}e^{-2h\tau}-C_{2}\right)e^{\frac{h\tau}{2}}\right]^{\frac{1}{3}},\label{eq: scale1}
\end{equation}
where $C_{1}$ and $C_{2}$ are the integration constants. Then the
Hubble parameter can be found as 
\begin{equation}
K\left(\tau\right)=\frac{\dot{a}}{a}=\frac{h\left(3C_{1}+C_{2}e^{2h\tau}\right)}{6\left(-C_{1}+C_{2}e^{2h\tau}\right)},\label{eq: hubble1}
\end{equation}
the effective equation of state, 
\begin{equation}
w\left(\tau\right)=-1-\frac{2\dot{K}}{3K^{2}}=-1+\frac{32C_{1}C_{2}e^{2h\tau}}{\left(3C_{1}+C_{2}e^{2h\tau}\right)^{2}}.\label{eq: eos1}
\end{equation}

\begin{figure}
\includegraphics[height=4cm]{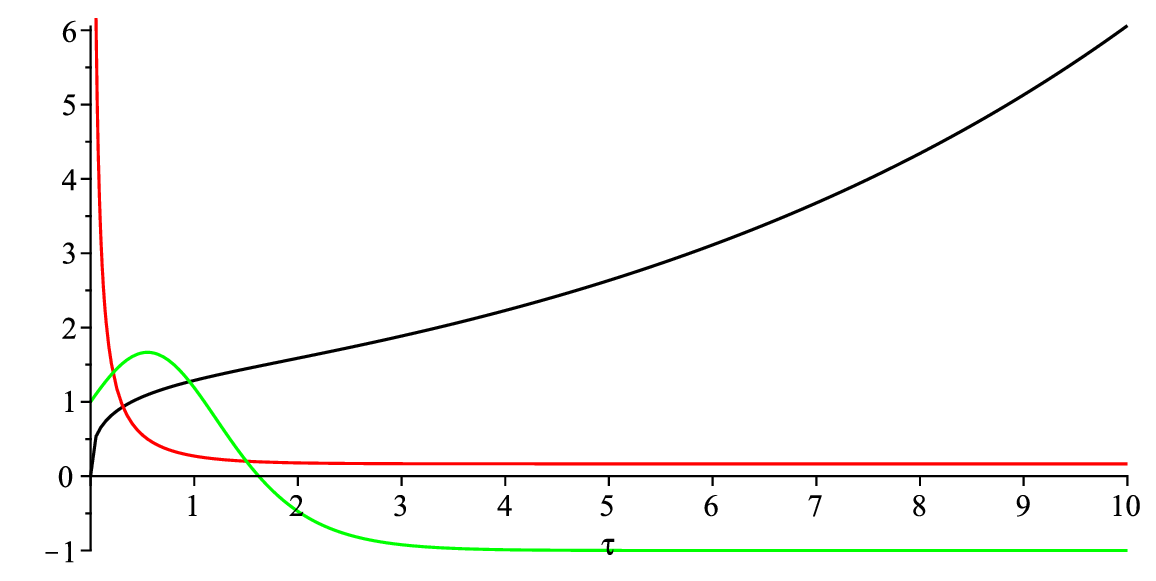}\includegraphics[height=4cm]{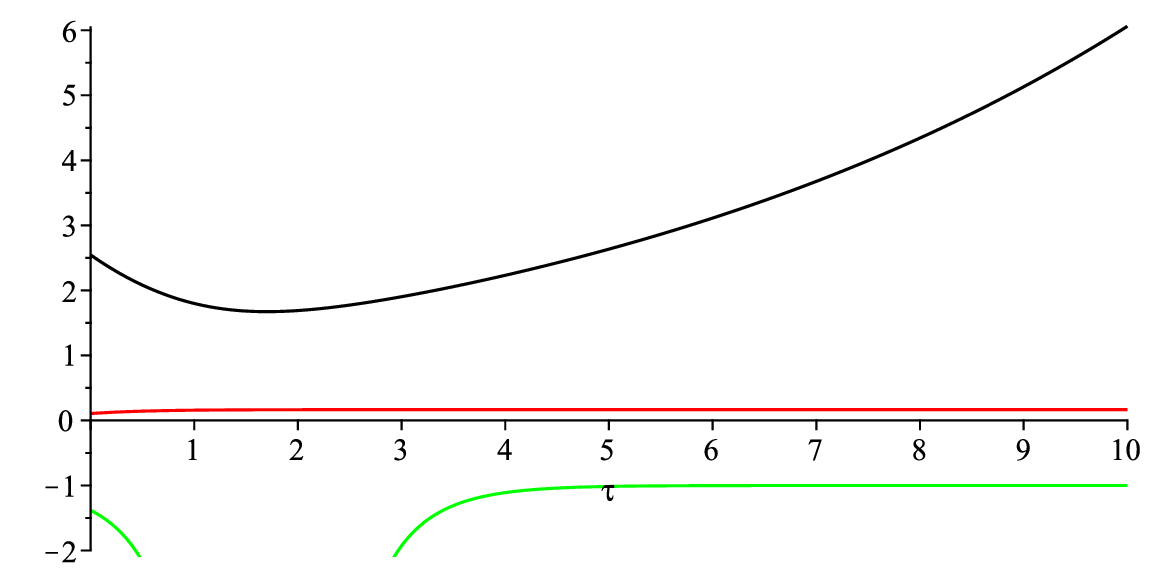}

\caption{\label{fig:The-scale-factor-1}The scale factor $a\left(\tau\right)$
(black line) in Eq.(\ref{eq: scale1}), the Hubble parameter $K\left(\tau\right)$
(red line) in Eq.(\ref{eq: hubble1}) and the effective equation of
state function $w\left(\tau\right)$ (green line) in Eq.(\ref{eq: eos1}).
The first graph with the parameters $h=1$, $C_{1}=-1$, $C_{2}=-1$
and the second with $h=1$, $C_{1}=10$, $C_{2}=-1$.}
\end{figure}

In Fig. \ref{fig:The-scale-factor-1}, we demonstrate the time evolution
of these parameters in two certain conditions. In both condition,
For the late time situations, the scale factor is increasing, the
analog of Hubble parameter for $\tau$ is almost remains constant
and the EoS function behaves like dark energy ($w\left(\tau\right)\approx-1$).

Moreover, by using Eqs.(\ref{eq: r(tau)_3}), (\ref{eq: r(tau)_def})
and (\ref{eq: scale1}) we get the following expressions 
\begin{equation}
r\left(\tau\right)=3C_{1}^{2}e^{h\tau},
\end{equation}
and 
\begin{equation}
f_{R}=\left(u_{0}-3\kappa C_{1}^{2}\right)e^{h\tau}-1,
\end{equation}
then the $f\left(R\right)$ function becomes 
\begin{equation}
f\left(R\right)=\left[\left(u_{0}-3\kappa C_{1}^{2}\right)e^{h\tau}-1\right]R+C_{3},
\end{equation}
and finally the Lagrange multiplier is found 
\begin{equation}
\gamma\left(\tau\right)=\frac{2\alpha\epsilon\left(e^{-h\tau}-u_{0}\right)-3C_{1}^{2}e^{h\tau}\left[3\kappa C_{1}^{2}\left(\alpha\epsilon R-2\right)-\alpha\epsilon u_{0}R+2u_{0}-2\right]+3C_{1}^{2}\left[-\alpha\epsilon\left(R-2\kappa-C_{3}\right)-2\lambda+2\right]}{3\alpha\epsilon C_{1}^{2}}.
\end{equation}

\subsubsection{Power-law evolution of $u\left(\tau\right)$}

\begin{figure}
\includegraphics[height=4cm]{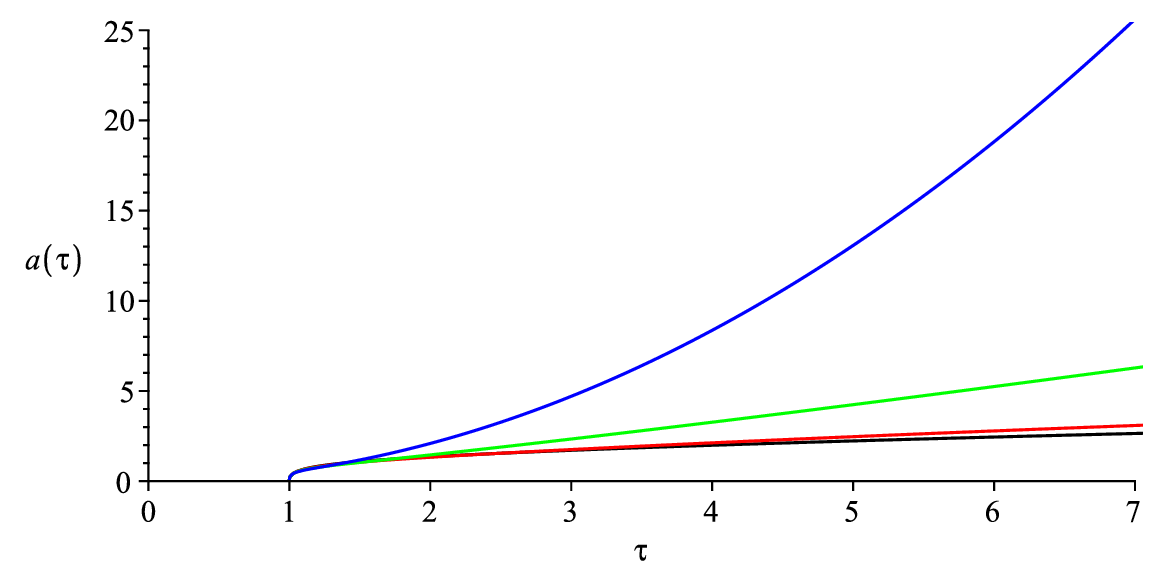}\includegraphics[height=4cm]{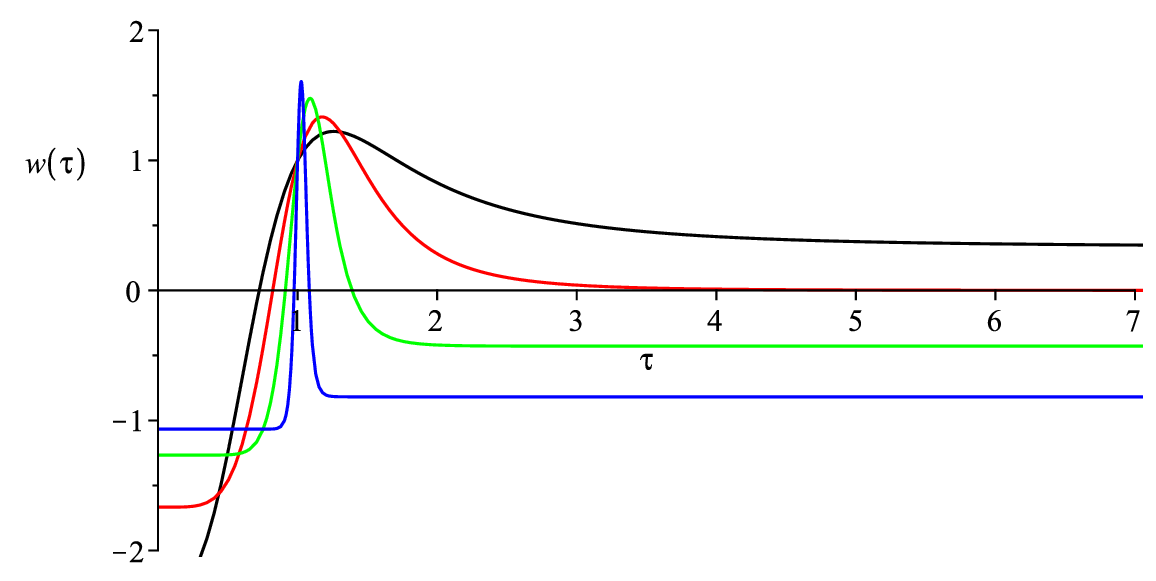}

\caption{\label{fig:The-scale-factor}The scale factor $a\left(\tau\right)$
(the first graph) (\ref{eq: scale2}), the effective equation of state
function $w\left(\tau\right)$ (the second graph) for the second case
(\ref{eq: eos2}). The parameters $C_{1}=-1$, $C_{2}=-1$ and $h=1$
(black line), $h=2$ (red line), $h=5$ (green line), $h=10$ (blue
line).\label{fig: u_exp-1}}
\end{figure}

Let us consider, the auxiliary metric function behave as a power-law
evolution $u\left(\tau\right)=\tau^{h}$. In this case, if we solve
Eq.(\ref{eq: u_a_equ}) with respect to the $a\left(\tau\right)$
then the scale factor becomes 
\begin{equation}
a\left(\tau\right)=\left[\frac{3\left(C_{1}\tau^{-\left(2h+1\right)}-C_{2}\right)}{\left(1+2h\right)\tau^{-\left(\frac{h}{2}+1\right)}}\right]^{\frac{1}{3}},\label{eq: scale2}
\end{equation}
and the Hubble parameter, 
\begin{equation}
K\left(\tau\right)=-\frac{3hC_{1}+C_{2}\left(h+2\right)\tau^{1+2h}}{6\tau\left(C_{1}-C_{2}\tau^{1+2h}\right)},\label{eq: hubble2}
\end{equation}
the effective equation of state, 
\begin{equation}
w\left(\tau\right)=-1+\frac{4\left[-8\left(h+\frac{5}{4}\right)C_{1}C_{2}h\tau^{1+2h}-2C_{2}^{2}\left(h+2\right)\tau^{4h+2}+3C_{1}^{2}h\right]}{9\tau^{h}\left(hC_{1}\tau^{-2}+\frac{\left(h+2\right)C_{2}\tau}{3}\right)^{2}}.\label{eq: eos2}
\end{equation}
In Fig. \ref{fig:The-scale-factor}, we present the time-dependence
of the scale factor and the effective EoS. In the given conditions,
for increasing $h$ parameters, the scale factor growing rapidly and
the EoS shows dark energy like behavior.

By using Eqs.(\ref{eq: r(tau)_3}), (\ref{eq: r(tau)_def}) and (\ref{eq: scale2})
we get the following expressions 
\begin{equation}
r\left(\tau\right)=3\tau^{h}C_{1}^{2},
\end{equation}
and 
\begin{equation}
f_{R}=\left(1-3\kappa C_{1}^{2}\right)\tau^{h}-1,
\end{equation}
then the $f\left(R\right)$ function becomes 
\begin{equation}
f\left(R\right)=\left[\left(1-3\kappa C_{1}^{2}\right)\tau^{h}-1\right]R+C_{3},
\end{equation}
and finally the Lagrange multiplier is found 
\begin{equation}
\gamma\left(\tau\right)=\frac{2\alpha\epsilon\left(\tau^{-h}-1\right)-3C_{1}^{2}\tau^{h}\left[3\kappa C_{1}^{2}\left(\alpha\epsilon R-2\right)-\alpha\epsilon R\right]+3C_{1}^{2}\left[-\alpha\epsilon\left(R-2\kappa-C_{3}\right)-2\lambda+2\right]}{3\alpha\epsilon C_{1}^{2}}.
\end{equation}

\subsection{The Parametric Friedmann equations}

Having observed that enforcing a conformal Ansatz, typically employed
in pure $f\left(R\right)$ theories, we will now demonstrate an alternative
solution to the connection equation. This approach avoids imposing
any constraints on the form of the function $f\left(R\right)$ that
defines the gravity Lagrangian. In order to examine the characteristics
of the cosmological solutions within the framework of the generalized
theory described by Eqs.(\ref{eq: eom_conn}) and (\ref{eq: eom_g-1}),
we adopt a method akin to the approach suggested in \citep{Makarenko:2014lxa}
for our specific model. Using the notation $\hat{q}$ and $\hat{q}^{-1}$
to denote $q_{\mu\nu}$ and $q^{\mu\nu}$, respectively. By defining
the object $\hat{\Omega}=\hat{g}^{-1}\hat{q}$, (\ref{eq: eom_g-1})
can be written as follows;

\begin{eqnarray}
|\hat{\Omega}|^{\frac{1}{2}}\hat{\Omega}^{-1}-\left\{ \left[\lambda-\frac{\alpha\epsilon}{2}\left(f\left(R\right)-\gamma\right)-\alpha f_{R}\right]\hat{I}+\alpha f_{R}\hat{\Omega}\right\}  & = & -\kappa^{2}\epsilon\hat{T},\label{eq: eom_omega}
\end{eqnarray}
where $\hat{I}$ is the identity matrix, and $\hat{T}$ denotes $T^{\mu\alpha}g_{\alpha\nu}$.
This equation establishes an algebraic relation between the object
$\hat{\Omega}$ and the matter. Now, multiplying this equation by
$\hat{\Omega}^{-1}$ and defining

\begin{equation}
\hat{B}=\frac{1}{2|\hat{\Omega}|^{\frac{1}{2}}}\left\{ \left[\lambda-\frac{\alpha\epsilon}{2}\left(f-\gamma\right)-\alpha f_{R}\right]\hat{I}-\kappa^{2}\epsilon\hat{T}\right\} ,
\end{equation}
we can write Eq.(\ref{eq: eom_omega}) in the more compact form 
\begin{equation}
\left(\hat{\Omega}^{-1}-\hat{B}\right)^{2}=\frac{\alpha f_{R}}{|\hat{\Omega}|^{\frac{1}{2}}}\hat{I}+\hat{B}^{2}.
\end{equation}

We can now consider the connection equation (\ref{eq: eom_conn}),
which can also be written as 
\begin{equation}
\nabla_{\beta}\left[\sqrt{-g}g^{\mu\nu}\Sigma_{\lambda}^{\,\,\nu}\right]=0,\label{eq: eom_conn_sigma}
\end{equation}
where we have defined 
\begin{equation}
\Sigma_{\lambda}^{\,\,\nu}=\left(|\hat{\Omega}|^{\frac{1}{2}}\left[\hat{\Omega}^{-1}\right]_{\lambda}^{\,\,\nu}+\alpha f_{R}\delta_{\lambda}^{\,\,\nu}\right).
\end{equation}
Assuming that $\Sigma_{\lambda}^{\,\,\nu}$ is invertible, as will
be the case of a perfect fluid to be considered in this work, we can
write the term within brackets in the above equation as $\sqrt{-g}\hat{g}^{-1}\hat{\Sigma}$
and look for an auxiliary metric $\hat{h}$ such that $\sqrt{-g}\hat{g}^{-1}\hat{\Sigma}=\sqrt{-h}\hat{h}^{-1}$thus
we can write 
\begin{equation}
\hat{h}=|\hat{\Sigma}|^{\frac{1}{2}}\hat{\Sigma}^{-1}\hat{g},\,\,\,\,\,\,\,\,\,\,\,\,\hat{h}^{-1}=\frac{1}{|\hat{\Sigma}|^{\frac{1}{2}}}\hat{g}^{-1}\hat{\Sigma}.\label{eq: metric_h}
\end{equation}
The connection equation (\ref{eq: eom_conn_sigma}) can thus be written
as $\nabla_{\beta}\left[\sqrt{-h}h^{\mu\nu}\right]=0$ which implies
that $\Gamma_{\mu\nu}^{\alpha}$ is the Levi- Civita connection of
$h_{\mu\nu}$.

Raising one index of this equation with $h^{\nu\alpha}$ and using
the definitions of $\hat{\Sigma}$ and $\hat{\Omega}$, we get 
\begin{equation}
R_{\mu}^{\,\,\beta}\left(h\right)=\frac{\Sigma_{\mu}^{\,\,\gamma}}{\epsilon|\hat{\Sigma}|^{\frac{1}{2}}}\left[\Omega_{\mu}^{\,\,\beta}-\delta_{\mu}^{\,\,\beta}\right].\label{eq: Ricci_h}
\end{equation}

\subsubsection{Perfect fluid scenarios}

For a perfect fluid with energy density $\rho$, pressure $P$, and
stress-energy tensor of the form 
\begin{equation}
T_{\mu\nu}=\left(\rho+P\right)u_{\mu}u_{\nu}+Pg_{\mu\nu},
\end{equation}
we find that

\begin{equation}
B_{\mu}^{\nu}=\frac{1}{2|\hat{\Omega}|^{\frac{1}{2}}}\left(\begin{array}{cc}
b_{1} & 0\\
0 & b_{2}\hat{I}_{3\times3}
\end{array}\right),
\end{equation}
where

\begin{equation}
b_{1}=\lambda-\alpha\left(\frac{\epsilon}{2}\left(f-\gamma\right)+f_{R}\right)+\kappa^{2}\epsilon\rho,
\end{equation}

\begin{equation}
b_{2}=\lambda-\alpha\left(\frac{\epsilon}{2}\left(f-\gamma\right)+f_{R}\right)-\kappa^{2}\epsilon P.
\end{equation}
By using these results one can find that

\begin{equation}
\Omega_{\mu}^{\,\nu}=2|\hat{\Omega}|^{\frac{1}{2}}\begin{pmatrix}w_{1} & 0\\
0 & w_{2}\hat{I}_{3x3}
\end{pmatrix},\label{eq: Omega1}
\end{equation}

\begin{equation}
\left(\hat{\Omega}^{-1}\right)_{\mu}^{\,\nu}=\frac{1}{2|\hat{\Omega}|^{\frac{1}{2}}}\begin{pmatrix}w_{1}^{-1} & 0\\
0 & w_{2}^{-1}\hat{I}_{3x3}
\end{pmatrix},\label{eq: Omega2}
\end{equation}
where 
\begin{equation}
w_{i}=\left[b_{i}+\sqrt{b_{i}^{2}+4\alpha f_{R}|\hat{\Omega}|^{\frac{1}{2}}}\right]^{-1}.
\end{equation}
In this case, if $f\left(R\right)$ term is negligible or the constant
$\alpha$ goes to zero we recover the result of the BI case. In this
scenario, if the $\alpha f_{R}$ term turns negative, the square root
has the potential to reach zero or even take on negative values, which
could lead to inconsistencies.

The determinant of (\ref{eq: Omega2}) leads to 
\begin{equation}
16|\hat{\Omega}|=\frac{1}{w_{1}w_{2}^{3}},\label{eq: Omega_det}
\end{equation}
whereas the trace of (\ref{eq: Omega1}) yields

\begin{equation}
4+\epsilon R=2|\hat{\Omega}|^{\frac{1}{2}}\left(w_{1}+3w_{2}\right).\label{eq: Omega_det_sqrt}
\end{equation}
By combining equations Eqs.(\ref{eq: Omega_det}) and (\ref{eq: Omega_det_sqrt}),
it should be theoretically possible to derive expressions for $R$
and $|\hat{\Omega}|$ based on the variables $\rho$ and $P$.

\subsubsection{General expressions for $\rho$ and $P$}

As mentioned earlier, equations (\ref{eq: Omega_det}) and (\ref{eq: Omega_det_sqrt})
establish algebraic relationships among the variables $\rho$, $P$,
$R$, and $|\hat{\Omega}|$ but only two of these variables are truly
independent. In this context, obtaining an expression for $\rho$
and $P$ in relation to $R$ and $|\hat{\Omega}|$ is quite straightforward,
and it doesn't necessitate specifying the specific $f\left(R\right)$
Lagrangian. This approach yields $\rho$ and $P$ in parametric form.
The idea is to start from (\ref{eq: Omega_det}) and transforming
it into the following format,

\begin{equation}
\frac{1}{b_{1}+\sqrt{b_{1}^{2}+4\alpha f_{R}|\hat{\Omega}|^{\frac{1}{2}}}}=\frac{b_{2}+\sqrt{b_{2}^{2}+4\alpha f_{R}|\hat{\Omega}|^{\frac{1}{2}}}}{16|\hat{\Omega}|}.\label{eq: Omega_b_relation}
\end{equation}
This relation can be inserted in (\ref{eq: Omega_det_sqrt}) to remove
the dependence on $\rho$ or to remove the dependence on $P$. For
instance, using (\ref{eq: Omega_b_relation}) to remove the dependence
on $\rho$ from (\ref{eq: Omega_det_sqrt}) and defining $\delta_{2}=b_{2}+\sqrt{b_{2}^{2}+4\alpha f_{R}|\hat{\Omega}|^{\frac{1}{2}}}$,
we get 
\begin{equation}
4+\epsilon R=2|\hat{\Omega}|^{\frac{1}{2}}\left(\frac{\delta_{2}^{3}}{16|\hat{\Omega}|}+\frac{3}{\delta_{2}}\right),\label{eq: Omega_det_sqrt-1}
\end{equation}
which allows us to write $P$ as 
\begin{equation}
\kappa^{2}\epsilon P=\lambda-\alpha\left(\frac{\epsilon}{2}\left(f-\gamma\right)+f_{R}\right)-\frac{\delta_{2}^{2}-4\alpha f_{R}|\hat{\Omega}|^{\frac{1}{2}}}{2\delta_{2}}.\label{eq: Pressure}
\end{equation}
A similar approach can be used to extract $\rho$ from

\begin{equation}
\delta_{1}=b_{1}+\sqrt{b_{1}^{2}+4\alpha f_{R}|\hat{\Omega}|^{\frac{1}{2}}},
\end{equation}
In this case, one gets

\begin{equation}
4+\epsilon R=2|\hat{\Omega}|^{\frac{1}{2}}\left(\frac{1}{\delta_{1}}+\frac{3\delta_{1}^{\frac{1}{3}}}{\left(4x\right)^{\frac{2}{3}}}\right),\label{eq: Omega_det_sqrt-1-1}
\end{equation}
The procedure is analogous to the previous case and yields

\begin{equation}
\kappa^{2}\epsilon\rho=-\left\{ \lambda-\alpha\left(\frac{\epsilon}{2}\left(f-\gamma\right)+f_{R}\right)-\frac{\delta_{2}^{2}-4\alpha f_{R}|\hat{\Omega}|^{\frac{1}{2}}}{2\delta_{2}}\right\} .\label{eq: EnergyDensity}
\end{equation}
As was already assumed at the beginning of this section, we focus
on a cosmological symmetry and use the unimodular metric in Eq.(\ref{eq: UnimodularMetric})
and use relations (\ref{eq: metric_h}) to find its relation with
the components of $h_{\mu\nu}$ necessary to use the field equations
(\ref{eq: Ricci_h}). Following a notation similar to that used in
\citep{olmo2012open}, we can write 
\begin{equation}
\Sigma_{\mu}^{\,\nu}=\begin{pmatrix}\sigma_{1} & 0\\
0 & \sigma_{2}\hat{I}_{3x3}
\end{pmatrix},\,\,\,\,\,\,\,\,\sigma_{i}=\alpha f_{R}+\frac{\delta_{i}}{2},\label{eq: Sigma}
\end{equation}
which implies

\begin{equation}
h_{tt}=-a\left(\tau\right)^{-6}\sqrt{\frac{\sigma_{2}^{3}}{\sigma_{1}}},
\end{equation}
\begin{equation}
h_{ij}=\sqrt{\sigma_{1}\sigma_{2}}a\left(\tau\right)^{2}\delta_{ij}=\Delta\left(\tau\right)a\left(\tau\right)^{2}\delta_{ij}.
\end{equation}
Recall that since $\sigma_{1}$ and $\sigma_{2}$ are functions of
$\rho$ and $P$, it follows that $\Delta$ is a function of time,
as we have explicitly written above. This is the only aspect we need
to know so far to proceed with the derivation of the Hubble equation.
After a bit of algebra, one gets

\begin{equation}
G_{\tau\tau}=3\left(K+\frac{\dot{\Delta}}{2\Delta}\right)^{2}.
\end{equation}
From the field equation (\ref{eq: Ricci_h}), we find that 
\begin{equation}
\epsilon G_{\tau\tau}=\frac{\sigma_{1}-3\sigma_{2}-2|\hat{\Omega}|^{\frac{1}{2}}\left(\sigma_{1}w_{1}-3\sigma_{2}w_{2}\right)}{2\sigma_{1}}.
\end{equation}
With this result, we can write 
\begin{equation}
\epsilon K^{2}=\frac{\sigma_{1}-3\sigma_{2}-2|\hat{\Omega}|^{\frac{1}{2}}\left(\sigma_{1}w_{1}-3\sigma_{2}w_{2}\right)}{2\sigma_{1}\left(1-\frac{3\left(1+\omega\right)\rho\Delta_{\rho}}{2\Delta}\right)^{2}}.\label{eq: Hubble}
\end{equation}

\subsubsection{A model $f\left(R\right)=R^{2}$}

To illustrate the procedure to deal with the theories presented in
this work, we consider a simple model characterized by a function
$f\left(R\right)=R^{2}$. In order to determine the impact of changing
the coefficient in front of the $R^{2}$ piece from the above action
on the Hubble function (\ref{eq: Hubble}), we need to work out the
dependence of $P$ and $\rho$ on $R$ and $|\hat{\Omega}|$ using
formulas (\ref{eq: Pressure}) and (\ref{eq: EnergyDensity}). The
first step is to solve $\delta_{2}$ from (\ref{eq: Omega_det_sqrt-1}).
To do so, it is convenient to introduce the redefinition $\delta_{2}=|\hat{\Omega}|^{\frac{1}{4}}$,
which turns (\ref{eq: Omega_det_sqrt-1}) into 
\begin{equation}
2z=\frac{x^{3}}{16}+\frac{3}{x},
\end{equation}
\begin{equation}
z=\frac{4+\epsilon R}{4|\hat{\Omega}|^{\frac{1}{4}}},\label{eq: relation_R_z}
\end{equation}

With this definition, one finds that (\ref{eq: Pressure}) can be
written as

\begin{equation}
\kappa^{2}\epsilon P=\lambda-\alpha\left(\frac{\epsilon}{2}\left(f-\gamma\right)+f_{R}\right)-\frac{|\hat{\Omega}|^{\frac{1}{4}}}{2}\frac{\left(x^{2}-4\alpha f_{R}\right)}{x}.\label{eq: Pressure-1}
\end{equation}
The equation for $\rho$ can be manipulated in a very similar way.
Introducing the replacement $\delta_{2}=\frac{16|\hat{\Omega}|^{\frac{1}{4}}}{y^{3}}$,
(\ref{eq: EnergyDensity}) becomes

\begin{equation}
2z=\frac{y^{3}}{16}+\frac{3}{y},
\end{equation}
which admits the same solution as $x$. Therefore, one finds that
(\ref{eq: EnergyDensity}) can be written as 
\begin{equation}
\kappa^{2}\epsilon\rho=-\left\{ \lambda-\alpha\left(\frac{\epsilon}{2}\left(f-\gamma\right)+f_{R}\right)-\frac{|\hat{\Omega}|^{\frac{1}{4}}}{8}\frac{\left(64-\alpha f_{R}y^{6}\right)}{y^{3}}\right\} .\label{eq: EnergyDensity-1}
\end{equation}
Thus, using Eqs.(\ref{eq: Pressure-1}) and (\ref{eq: EnergyDensity-1}),
we can write 
\begin{equation}
|\hat{\Omega}|^{\frac{1}{4}}=\frac{2\left(1+\omega\right)\left[\lambda-\alpha\left(\frac{\epsilon}{2}\left(f-\gamma\right)+f_{R}\right)\right]}{\frac{\left(x^{2}-4\alpha f_{R}\right)}{x}+\frac{\omega}{4}\frac{\left(64-\alpha f_{R}y^{6}\right)}{y^{3}}}.
\end{equation}
Taking into account of Eq.(\ref{eq: relation_R_z}), this equation
establishes a relation between $R$ and $z$ which is also allow us
to show $|\hat{\Omega}|$ as a function of $z$. To illustrate this
point, consider the case $\alpha f\left(R\right)=-a\epsilon R^{2}/4$
which satisfies the conditions the BI theory ($a\rightarrow0$) and
the BI-$f\left(R\right)$ case without the $R^{2}$ term ($a\rightarrow1$).
For an arbitrary $\omega$ can be found such as $\omega=0$, the curvature
function in terms of $z$ can be found as, 
\begin{equation}
\epsilon R\left(z\right)=\frac{x^{2}+a\left(8-4xy\right)\pm\sqrt{16a^{2}\left(xz-2\right)^{2}+8ax\left(xz-2\right)^{2}\left(x-2z\left(\alpha\epsilon\gamma+2\right)\right)+x^{4}}}{2a\left(xz-2\right)},
\end{equation}
which is valid for any $a\neq0$.

\section{Conclusion}

\label{sec-conclusion}

In this paper, we propose a novel modified gravity theory, namely
the extension of Born-Infeld-$f(R)$ (BI-$f(R)$) gravity, in the
context of unimodular gravity. In particular, we first formulate the
action corresponding to the generalized BI-$f(R)$ gravity, and then,
we present a reconstruction scheme of this unimodular extension to
achieve various cosmological eras of the universe.

Based on the fact that the determinant of the space-time metric is
constrained to be unity in unimodular gravity, we have investigated
which metric proves to be compatible with such a constraint since
the standard FRW metric is unable to do so. Similar to the standard
unimodular theory, it turns out that the constraint on the determinant
of the metric can be imposed in the present context by introducing
a Lagrange multiplier in the gravitational action. Having formulated
the action of the unimodular BI-$f(R)$ gravity (that indeed respects
the constraint of the metric), we derive the equations of motion by
employing the metric formalism. The resulting equations immediately
propose a reconstruction method, by which, one can determine the form
of unimodular BI-$f(R)$ gravity corresponding to some given cosmological
evolution. It further opens the avenue to find which cosmological
evolution corresponds to a given unimodular BI-$f(R)$ gravity. The
resulting cosmological scenario seems to be different from the ordinary
$f(R)$ gravity, as was probably expected. Interestingly, the unimodular
generalization of BI-$f(R)$ gravity turns out to be suitable for
inflation (de-Sitter and quasi de-Sitter) along with power law cosmology
which actually includes from reheating to the Standard Big Bang cosmology
(depending on the exponent of the power law scale factor). For quasi
de-Sitter inflation, we consider the Starobinsky kind of inflation
which has an exit, and also, the observable indices (like the spectral
index and the tensor-to-scalar ratio) are in agreement with the Planck
data. Here it deserves mentioning that in the case of the Starobinsky
inflation, the form of $f(R)$ in the unimodular BI-$f(R)$ theory
is different than that of the standard $f(R)$ theory. The potential
applications of the reconstruction method we present in this paper
are quite many since it is possible to realize various cosmological
evolutions, which are exotic for the standard Einstein-Hilbert gravity
and cannot be realized in that case, for example bouncing cosmologies.
In this regard, we show that apart from the inflation, such generalized
BI-$f(R)$ theory is also capable to trigger non-singular bouncing
cosmology for suitable forms of the unimodular Lagrange multiplier.
Thus as a whole, we determine the required form of $f(R)$ and the
unimodular Lagrange multiplier to achieve inflation and the Standard
Big Bang evolution of the universe.

However, we would like to mention that we have not described the universe's
evolution as a unified picture. It will be interesting to address
the unification of various cosmological eras in the context of unimodular
BI-$f(R)$ gravity, which we expect to study elsewhere. Also, an issue
we did not address in this paper is the cosmological behaviour of
BI-$f(R)$ near mild singularities, like the Type IV singularity.
Therefore, a concrete analysis of the unimodular BI-$f(R)$ gravity
theory near mild and even crushing types singularities is of some
interest and should be appropriately addressed in a future work (for
more details about the cosmological singularities, see recent reviews
in \citep{Haro:2023Finite-Time,Trivedi:2023RecentAdvances}). Moreover,
it is interesting that there exists a mimetic extension of unimodular
$f(R)$ gravity \citep{Nojiri:2016ppu}, and by following this approach,
one can extend the current theory to mimetic unimodular BI-$f(R)$
gravity. It would be of interest to study this question in case of
the gravity model under consideration. We also note that the recent
review \citep{Vagnozzi:2023SevenHints} is devoted to the study of
modifications of early and late universe descriptions to solve the
current cosmic tensions. This will be done elsewhere. We hope to address
some of the aforementioned issues in a future work.

\section*{Declaration of competing interest }

The authors declare no conflicts of interest.

\section*{Data Availability Statement}

No data was used for the research described in the article. 

\section*{Acknowledgments}

This work was partially supported by MICINN (Spain), project PID2019-104397GB-I00
and by the program Unidad de Excelencia Maria de Maeztu CEX2020-001058-M,
Spain (S.D.O).

\bibliography{BIFR_Unimodular}

\end{document}